\documentclass{PoS}
\usepackage{amssymb,amsmath}

\title{Excited-Nucleon Spectroscopy with 2+1 Fermion Flavors}

\ShortTitle{Excited-Nucleon Spectroscopy with 2+1 Fermion Flavors}

\author{Saul Cohen \\Department of Physics, Boston University, Boston, MA 02215
}
\author{John M. Bulava, Justin Foley, Colin Morningstar and Ricky Wong\\Department of Physics, Carnegie Mellon University, Pittsburgh, PA 15213}

\author{Robert G. Edwards, B\'alint Jo\'o, David G. Richards\\Thomas Jefferson National
Accelerator Facility, Newport News, VA 23606}

\author{Eric Engelson and Stephen J. Wallace\\Department of Physics, University of Maryland, College Park, MD 20742, USA}

\author{K. Jimmy Juge\\Department of Physics, University of the Pacific, Stockton, CA 95211, USA}

\author{\speaker{Huey-Wen Lin\footnote{HWL is supported by the U.S. Dept. of Energy under Grant No. DE-FG03-97ER4014 and DE-AC05-06OR23177.}} \\
Department of Physics, University of Washington, Seattle, WA 98195-1560 \\
        E-mail: \email{hwlin@phys.washington.edu}}
        
\author{Nilmani Mathur\\Department of Theoretical Physics, Tata Institute of Fundamental Research, India}

\author{Michael J. Peardon and Sin\'ead M. Ryan\\School of Mathematics, Trinity College, Dublin 2, Ireland}

\abstract{We present progress made by the Hadron Spectrum Collaboration (HSC) in determining the tower of excited nucleon states using 2+1-flavor anisotropic clover lattices. The HSC has been investigating interpolating operators projected into irreducible representations of the cubic group in order to better calculate two-point correlators for nucleon spectroscopy; results are published for quenched and 2-flavor anisotropic Wilson lattices. In this work, we present the latest results using a new technique, distillation, which allows us to reach higher statistics than before. Future directions will be outlined at the end.}

\FullConference{The XXVII International Symposium on Lattice Field Theory\\
		 July 26-31, 2009\\
		 Peking University, Beijing, China}

\begin{document}

\section{Introduction}
\vspace{-0.3cm}
A new generation of experiments devoted to hadron spectroscopy, including GlueX at Jefferson Lab, PANDA at GSI/FAIR and BES~III intend to make measurements with unprecedented precision and in previously unexplored mass ranges and quantum numbers.
Both in meson and baryon spectroscopy, there are many experimentally observed excited states whose physical properties are poorly understood and could use theoretical input from lattice QCD to solidify their identification. Aside from masses, other excited-state quantities that could be computed on the lattice, such as form factors and coupling constants, would be useful to groups such as the Excited Baryon Analysis Center (EBAC) at Jefferson Lab, where dynamical reaction models have been developed to interpret experimentally observed properties of excited nucleons in terms of QCD~\cite{ebac}. 
In certain cases, input from the lattice may be helpful in determining the composition of controversial states, which may be interpreted as ordinary hadrons, tetra- or pentaquarks, hadronic molecules or unbound resonances.

Among the excited nucleon states, the nature of the Roper resonance, $N(1440)\ P_{11}$, has been the subject of interest since its discovery in the 1960s. It is quite surprising that the rest energy of the first excited state of the nucleon is less than the ground-state energy of the nucleon's negative-parity partner, the $N(1535)\ S_{11}$~\cite{Yao:2006px}, a phenomenon never observed in meson systems. There are several interpretations of the Roper state, for example, as the hybrid state that couples predominantly to QCD currents with some gluonic contribution 
or as a five-quark (meson-baryon) state~\cite{others}

\begin{figure}[t]
\begin{center}
\includegraphics[width=0.32\textwidth]{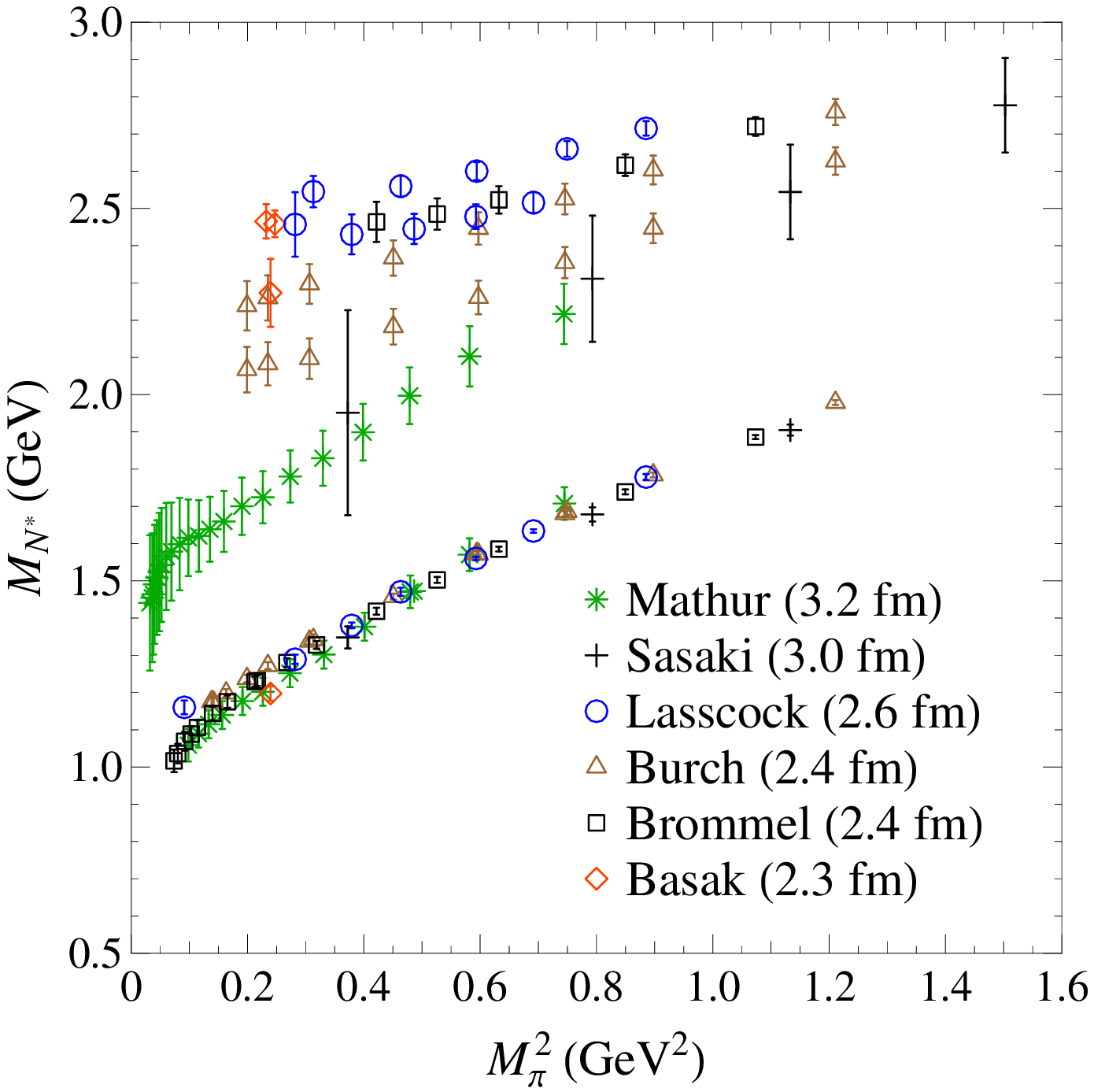} 
\includegraphics[width=0.32\textwidth]{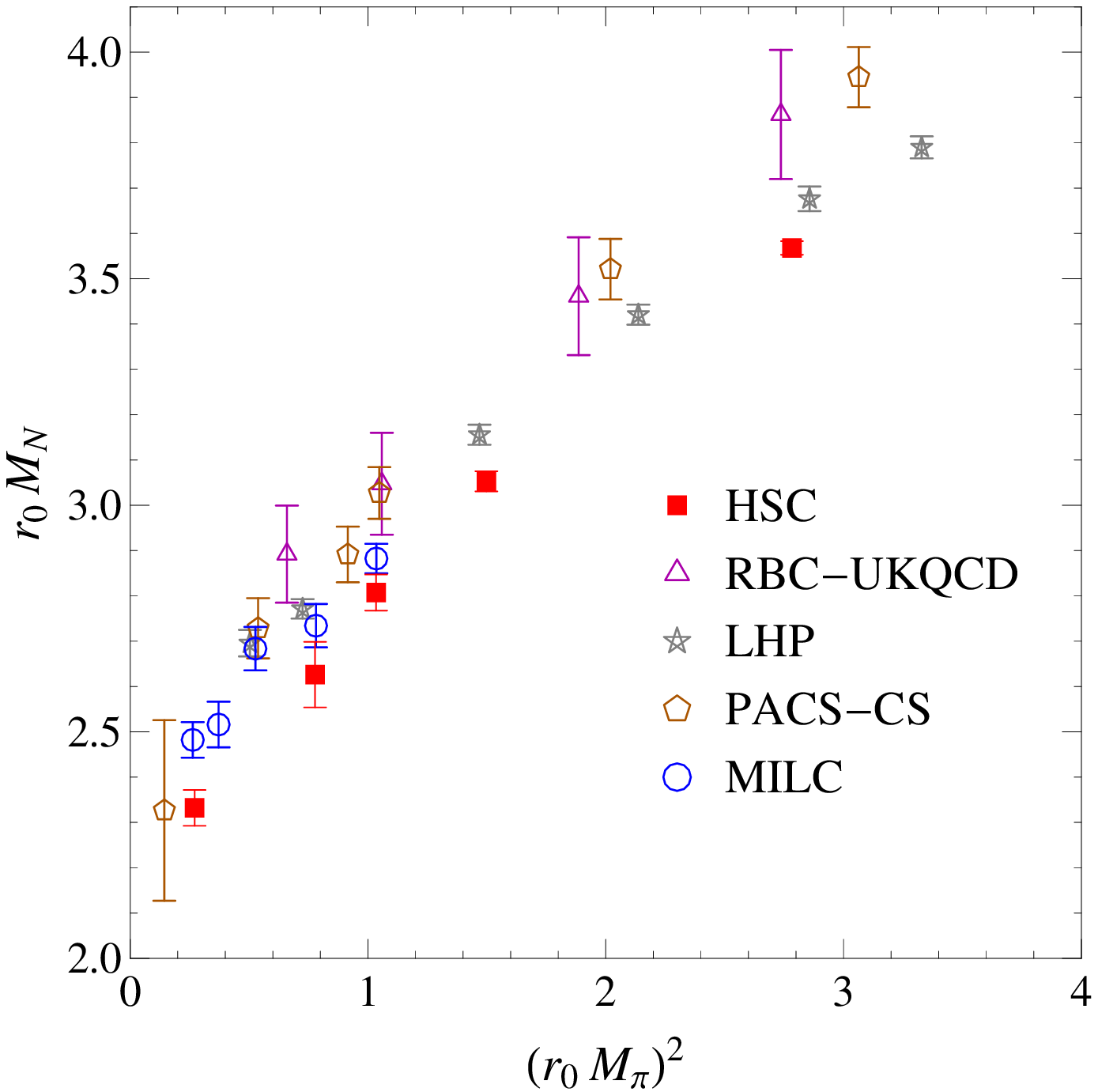} 
\includegraphics[width=0.32\textwidth]{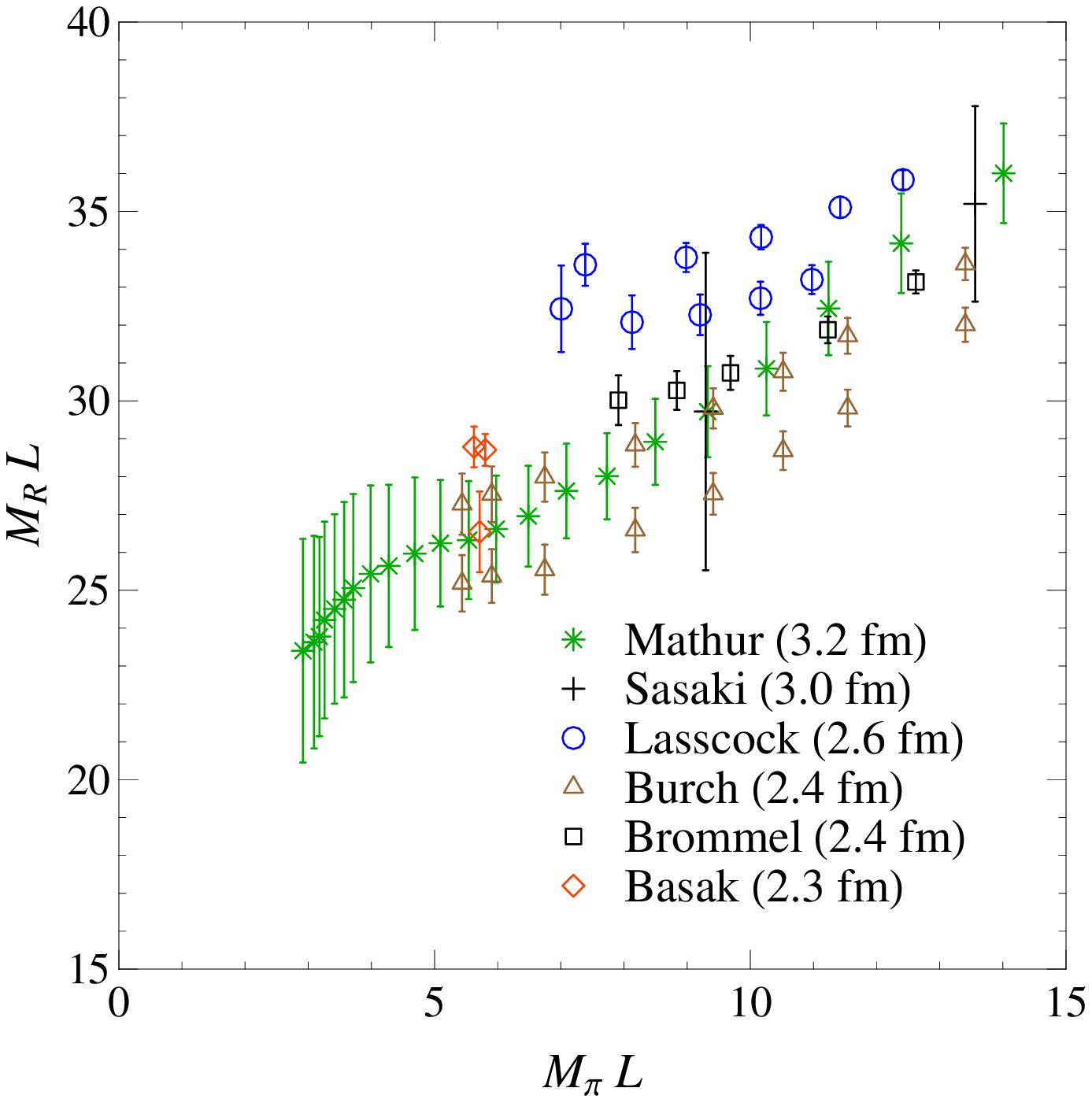} 
\end{center}
\vspace{-0.3in}
\caption{\label{fig:Resonance-FV} Summary of published quenched ($N_f=0$) lattice QCD calculations of the nucleon and Roper masses in GeV (left) and in terms of the dimensionless product of the Roper mass and lattice size $L$ (right). The middle panel shows the dimensionless products of lattice nucleon mass and $r_0$ as a function of $m_\pi r_0$ from $N_f=2+1$ ensemble.
}
\end{figure}

Due to the greater impact of systematic errors, such as finite-volume effects and discretization errors, on excited states, there is an essential need to examine calculations of excited masses more carefully during analysis. It is common in such calculations to use the ground-state masses, such as nucleon mass, as a starting point for analysis but without further checks of the consistency of the approach. The nucleon mass has been demonstrated with good consistency (a great demonstration of the universality of lattice QCD using different fermion actions by various groups with independent analyses) in both quenched and dynamical $N_f=2+1$ cases. For dynamical ensembles, there are more variables in terms of algorithm, scale setting, etc., but an approximate universality is achieved among different groups (see middle panel of Fig.~\ref{fig:Resonance-FV}). However, beyond the ground state, there exists a diverse distribution of excited-state masses as functions of $m_\pi$.
There are big discrepancies in the calculated nucleon first-excited mass, creating an apparently chaotic atmosphere, as shown in the left panel of Fig.~\ref{fig:Resonance-FV}.
Note that the errorbars here are just statistical, none of the systematic errors (such as quenching, finite-volume effect, etc.) are estimated. We re-address the same issue by modifying the axes to be in terms of the dimensionless quantity $ML$
(as shown in right panel of Fig.~\ref{fig:Resonance-FV});
we find that the Roper masses are roughly inversely proportional to lattice size.
Now we see a better agreement (or universality) among the lattice QCD Roper-mass calculations; most of the Roper masses agree within 2 standard deviations of the numbers in Ref.~\cite{Lasscock:2007ce}. This suggests that finite-volume effects can be more severe for excited states than ground states and that careful examination of such systematic errors is crucial.
Similarly, we cannot ignore other systematic errors that may arise even if we did not observe the effect in the ground state. Excited-state analyses should proceed with greater caution.

To tackle the challenge of extracting reliable excited-state energies, the Hadron Spectrum Collaboration (HSC) has been devoting resources and effort into resolving the mystery from the fundamental point of view. It is difficult to extract the excited state reliably without sufficient information carried in the nucleon correlators.
We need correlators which clearly contain the masses of specific quantum numbers and allow us access to data points that reach higher excited states before the signal decays exponentially away.
To successfully and reliably extract these excited-state energies, we need better resolution in temporal direction (large superfine isotropic lattices or anisotropic lattices), separation of the signals for individual states (variational method), and most importantly, operators that have good overlap with various desired quantum numbers (cubic-group irreducible representations and some way to put in many operators that contribute unique signals).

\section{ $N_f=2+1$ Clover Anisotropic Lattices and Updates}
\vspace{-0.3cm}
To improve our ability to extract higher excited energies (even with the aid of the variational method), we need to use finer temporal lattice spacings. This allows us to use as many time slices of information as possible to reconstruct more accurately the signal due to a particular state, since the signal in the Euclidean space declines exponentially as $E t$. In principle, we could use very fine lattices while keeping the volume big enough to avoid the ``squeezing'' systematic effect; however, the computational cost scales significantly (power of 5--6) with the lattice discretization. We adopt the approach of anisotropic clover lattices to keep the spatial lattice spacing coarse, avoiding finite-volume systematic error, and to make the temporal lattice spacing fine enough to extract towers of nucleon excited states. However, tuning the $O(a)$-improved parameters for fermion actions in the dynamical gauge generation correctly is a much more difficult task for anisotropic lattices. We use Symanzik-improved gauge action and clover fermion action with 3-dimensional stout-link smeared gauge fields; the gauge ensemble is generated using the (R)HMC algorithm. Our spatial lattice spacing is $a_s=0.1227(8)$~fm (determined using $m_\Omega$), and the renormalized anisotropy $\xi_R=a_s/a_t$ is $3.5$.
Ref.~\cite{Edwards:2008ja} shows a detailed study of the dynamical anisotropic lattice parameter settings, and Ref.~\cite{Lin:2008pr} reports basic lattice properties along with
the ground-state hadron spectrum.

Since then, we have also moved on to generate $m_\pi=230$ MeV $24^3$ and $32^3$ lattices. A stream of  $m_\pi=180$ MeV ensembles is slowly progressing (for both $24^3$ and $32^3$ volumes). In Ref.~\cite{Lin:2008pr}, we demonstrated and compared various strange-quark mass-setting approaches. We tuned the parameter $s_\Omega=\frac{9(2m_K^2-m_\pi^2)}{4m_\Omega^2}$ as close as possible to the corresponding experimental value at the $SU(3)_f$-symmetric point (in our case, it is the 875-MeV--pion ensemble). When we reduce the sea-quark mass, the $s_\Omega$ parameter remains roughly constant as low as the 383~MeV pion ensemble. (Furthermore, the $s_\Omega$ parameter is more sensitive to the strange-quark mass than other methods, such as $J$-parameter. Since it is a ratio, there is no need to worry about estimating the shift in the lattice spacing when decreasing the sea-quark mass or generating more statistics.) We measure the same quantity on the 230-MeV $24^3$ ensemble and $s_\Omega$ does not deviate from the expected value. The $s_\Omega$ parameter is a stable and useful observable for setting the strange-quark mass; we have not observed any notable deviation for a wide range of sea-quark masses.

\begin{figure}[t]
\begin{center}
\includegraphics[width=0.32\textwidth]{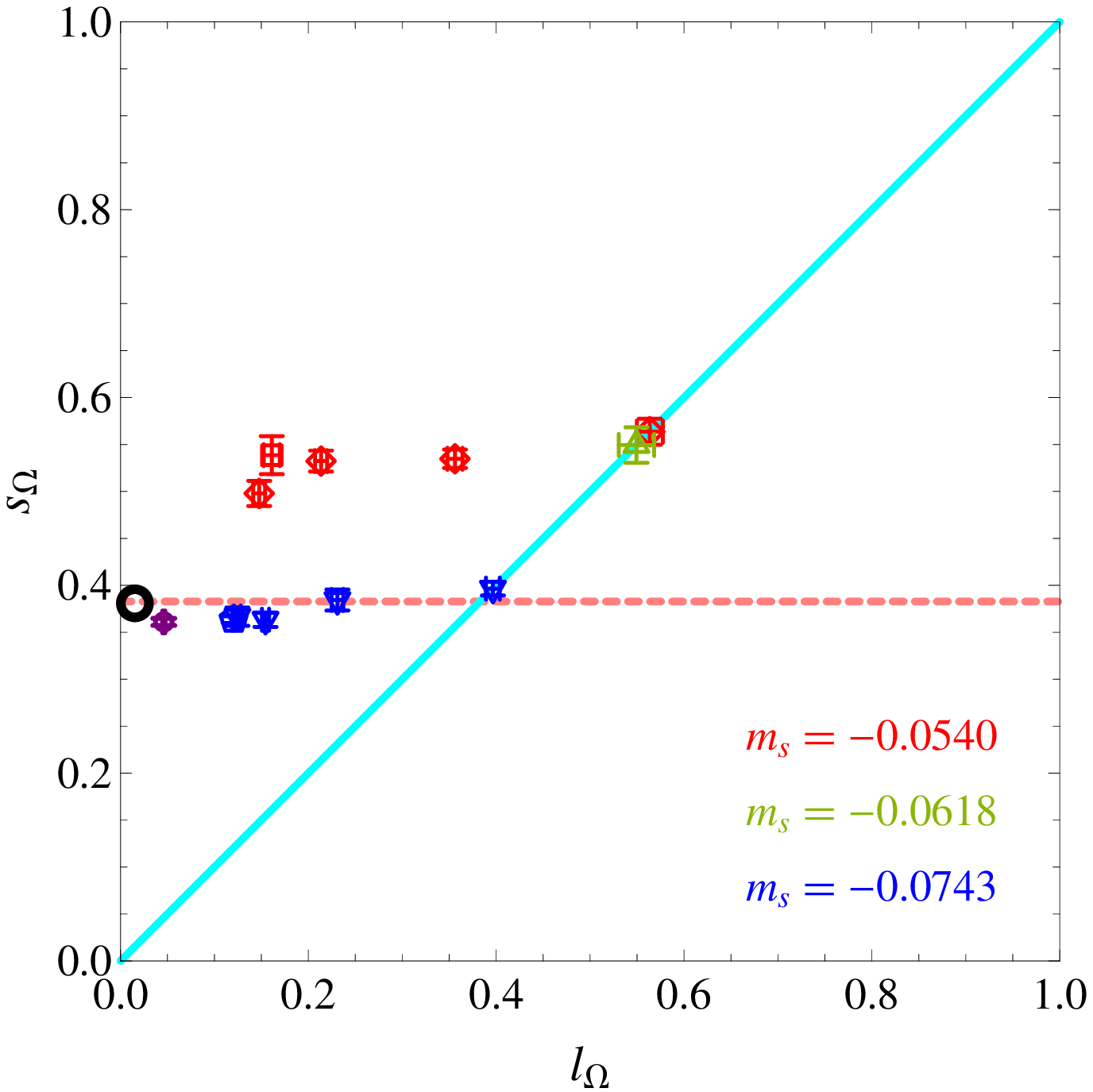}
\includegraphics[width=0.32\textwidth]{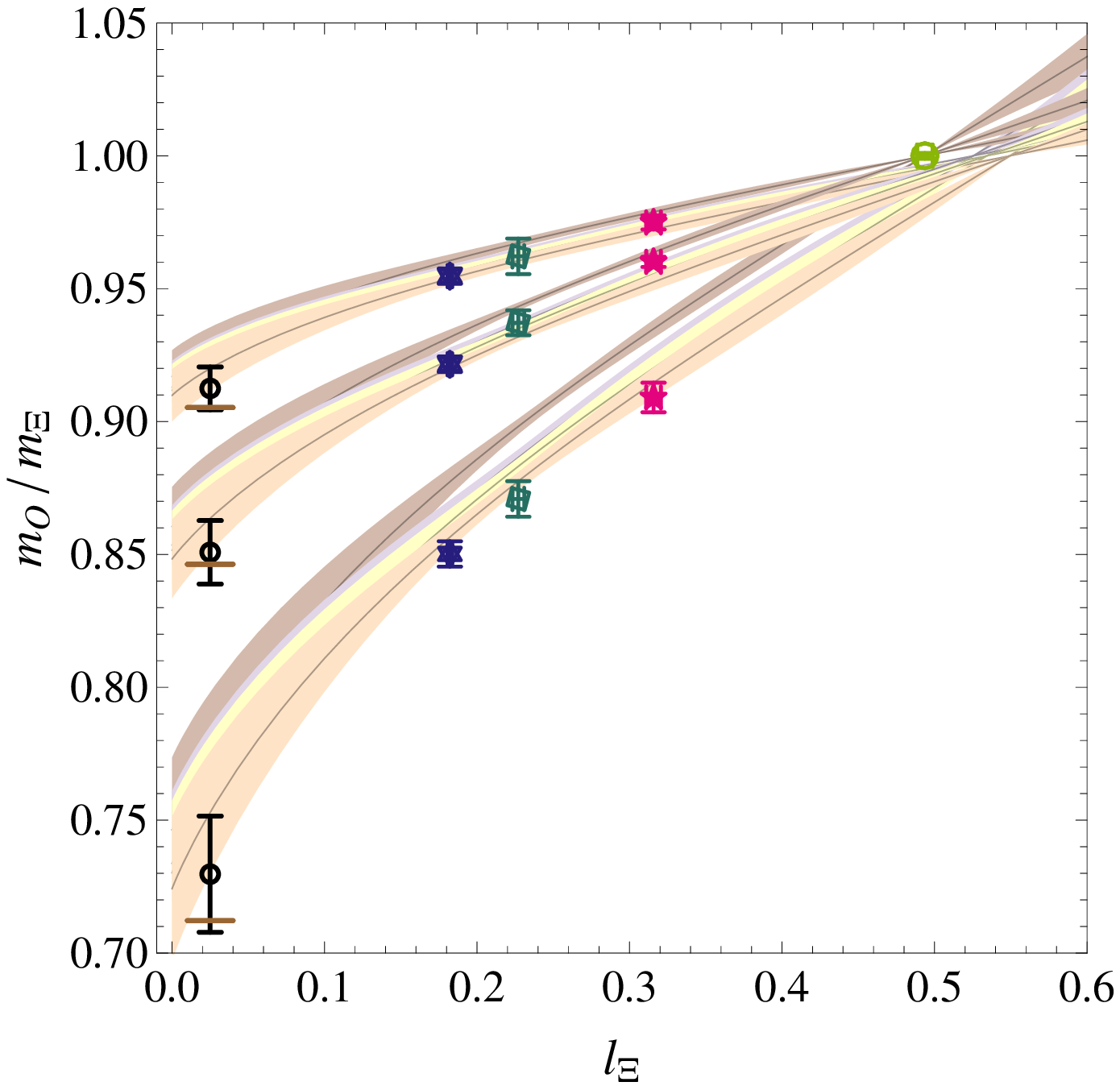}
\includegraphics[width=0.32\textwidth]{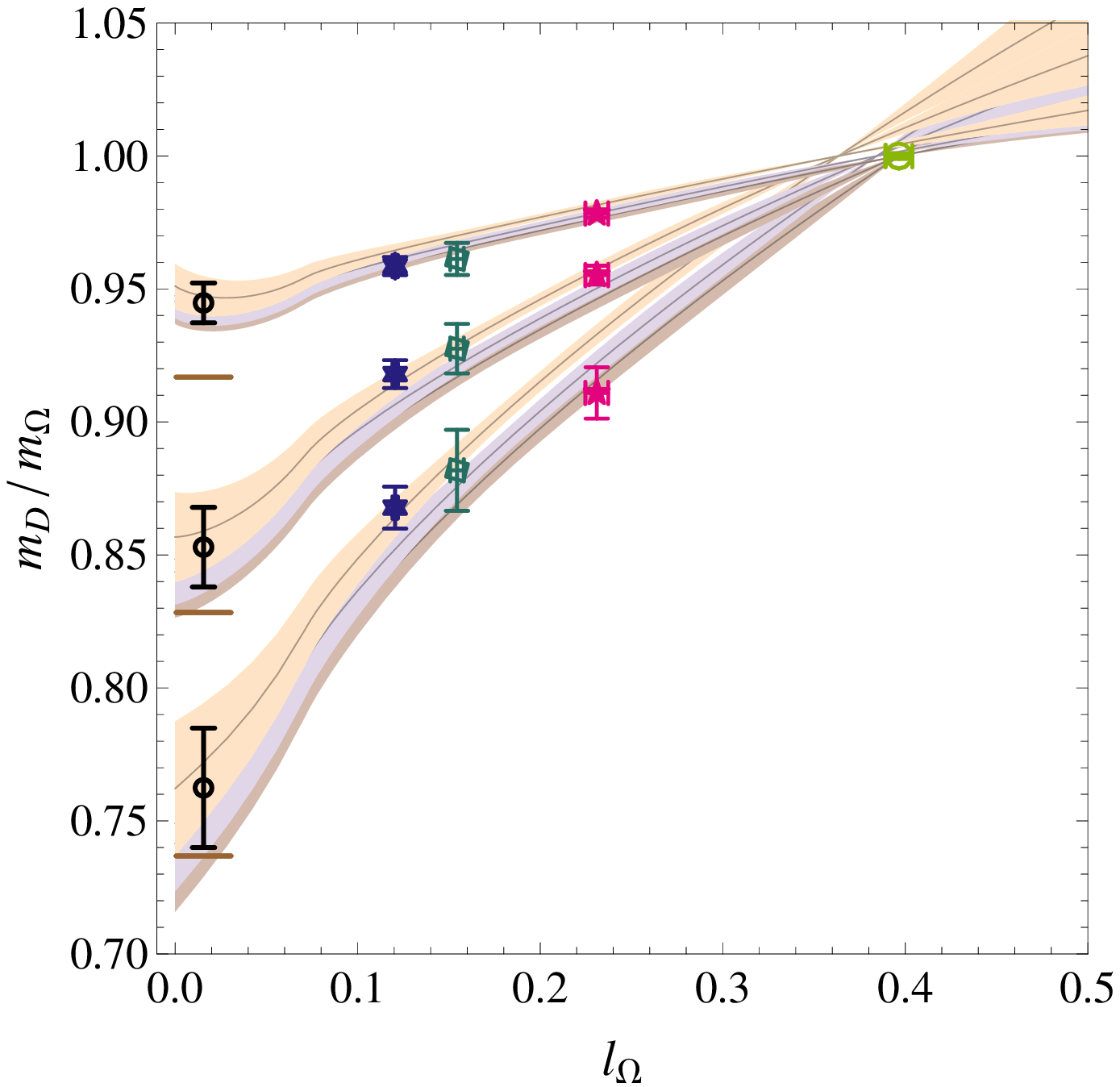}
\end{center}
\vspace{-0.3in}
\caption{\label{fig:ChPTMassRatio}
(Left panel) The location of the dynamical ensembles used in this work in the $s_\Omega$-$l_\Omega$ plane. The leftmost circle (black) indicates the physical point \{$l_\Omega^{\rm phys}$, $s_\Omega^{\rm phys}$\}, while the red, green and blue points are the lattices with $a_tm_s=-0.0540, -0.0618, -0.0743$ respectively in Ref.~\cite{Lin:2008pr}. The horizontal dashed (pink) line indicates constant $s_\Omega$ at the physical point, and the diagonal line indicates three-flavor degenerate theories. The purple points are from recent measurements of the 230~MeV ensemble, showing that it does not deviate away from the $N_f=3$ value.
Mass-ratio chiral extrapolations as functions of the $l_\Xi$ and $s_\Xi$ for octets (center  panel) and $l_\Omega$ and $s_\Omega$ for decuplets (right panel).
The lines indicate the ``projected'' leading chiral extrapolation fits in $l_\Xi$ and $s_\Xi$ ($l_\Omega$ and $s_\Omega$) while keeping the other one fixed. The black (circular) point is the extrapolated point at physical $l_\Xi$ and $s_\Xi$ ($l_\Omega$ and $s_\Omega$).
The masses for $N$, $\Sigma$ and $\Lambda$ are 0.962(29), 1.203(11), 1.122(16)~GeV, respectively, and the masses for $\Delta$, $\Sigma^*$ and $\Xi^*$ are 1.275(38), 1.426(25) and 1.580(13)~GeV, respectively.
}
\end{figure}

In Ref.~\cite{Lin:2008pr}, we reported a naive extrapolation in mass ratios (using $m_\Omega$ or $m_\Xi$ as the reference mass) to obtain the meson and baryon masses. Note that at each sea-strange mass, we only use the larger-volume set for the lightest pion mass to avoid large finite-volume effects. The advantage of using the mass ratios instead of the masses is that we gain smaller statistical (and possibly systematic) errors than the mass itself due to cancellation of systematics and removal of the ambiguity associated with setting the lattice spacing. Using the same data, we update the mass-ratio chiral extrapolations by modifying the next-to-leading-order heavy-baryon chiral perturbation theory. We find the finite-volume corrections are negligible according to ChPT estimates and the extrapolated baryon masses agree with experimental values significantly better than with a linear extrapolation. Preliminary results are shown in Fig.~\ref{fig:ChPTMassRatio}.

\section{Methodology: Cubic-Group Operators and Distillation}
\vspace{-0.2cm}
The Hadron Spectrum Collaboration (HSC) has been investigating interpolating operators projected into irreducible representations (irreps) of the cubic group~\cite{HSC2005}
in order to better calculate two-point correlators for nucleon spectroscopy.
Baryon correlation functions were evaluated using the displaced-quark operators described in Refs.~\cite{HSC2005} 
and employed in spectrum studies in Refs.~\cite{HSCnum}. 
In the cubic group $O_h$, for baryons, there are four two-dimensional irreps $G_{1g}, G_{1u}, G_{2g}$, $G_{2u}$ and two four-dimensional irreps $H_g$ and $H_u$. (The subscripts ``$g$'' and ``$u$'' indicate positive and negative parity, respectively.) Each lattice irrep contains parts of many continuum states. The $G_1$ irrep contains $J=\frac{1}{2},\frac{7}{2},\frac{9}{2},\frac{11}{2},\dots$ states, the $H$ irrep contains $J=\frac{3}{2},\frac{5}{2},\frac{7}{2}, \frac{9}{2},\dots$ states, and the $G_2$ irrep contains $J=\frac{5}{2},\frac{7}{2},\frac{11}{2},\dots$ states. The continuum-limit spins $J$ of lattice states must be deduced by examining degeneracy patterns among the different $O_h$ irreps.
Using these operators, we construct an $r \times r$ correlator matrix and extract individual excited-state energies by fitting with single- and multiple-exponential functions.

However, the calculations using these cubic-group operators require multiple orientations in order to maximally overlap with a wide range of quantum numbers, which is quite expensive. Furthermore, as we go to lighter and lighter pion masses, there will be more decay modes open, even for the lowest energy at a specific quantum number. We need to extend the matrix to include the multiple-particle operators (so that we can further understand the nature of the ``state'' in our calculation) and ``disconnected'' operators. Further, we need to achieve better precision for each state to distinguish among them.

A new way to calculate timeslice-to-all propagators, ``distillation'', has been proposed in Ref.~\cite{Peardon:2009gh}. The method is useful for creating complex operators, such as those used in the variational method, allows the operators to be decided after performing the Dirac inversions and reduces the amount of time needed for contractions. Distillation uses color-eigenvector sources to improve on noisy estimators, giving better coverage of relevant degrees of freedom. Increasing the number of ``hits'' improves statistics faster than 1/$\sqrt{N}$. The method can combine with stochastic methods, which might be desirable if the number of sources needed to cover the volume becomes too large.

The distillation operator on time-slice $t$ can be written as
\begin{equation}
 \Box(t) = V(t) V^\dagger(t)
 \rightarrow
\Box_{xy}(t) = \sum_{k=1}^N v_x^{(k)} (t) v_y^{(k)\dag} (t),
    \label{eqn:box}
\end{equation}
where the $V(t)$ is a matrix containing the first through $k^{\rm th}$ eigenvectors of the lattice spatial Laplacian.
The baryon operators involve displacements (${\cal D}_i$) as well as coefficients ($S_{\alpha_1\alpha_2\alpha_3}$) in spin space:
\begin{equation}
\chi_B(t) = \epsilon^{abc} S_{\alpha_1\alpha_2\alpha_3}
({\cal D}_1\Box d)^a_{\alpha_1}
({\cal D}_2\Box u)^b_{\alpha_2}
({\cal D}_3\Box u)^c_{\alpha_3}(t),
\end{equation}
where the color indices of the quark fields acted upon by the displacement operators are contracted with the antisymmetric tensor, and sum over spin indices. Then one can construct the two-point correlator; for example, in the case of proton,
\begin{align}
C^{(2)}_B[\tau_d,\tau_u,\tau_u](t',t) &=
  \Phi^{(i,j,k)}(t')
  \tau_d^{(i,\bar{i})}(t',t)
  \tau_u^{(j,\bar{j})}(t',t)
  \tau_u^{(k,\bar{k})}(t',t)
  \Phi^{(\bar{i},\bar{j},\bar{k})*}(t)
  \nonumber\\
 &\quad- \Phi^{(i,j,k)}(t')
  \tau_d^{(i,\bar{i})}(t',t)
  \tau_u^{(j,\bar{k})}(t',t)
  \tau_u^{(k,\bar{j})}(t',t)
  \Phi^{(\bar{i},\bar{j},\bar{k})*}(t),
\end{align}
where the ``baryon elemental''
\begin{equation}
 \Phi^{(i,j,k)}_{\alpha_1\alpha_2\alpha_3}(t) = \epsilon^{abc}
\left({\cal D}_1 v^{(i)}\right)^{a}
\left({\cal D}_2 v^{(j)}\right)^{b}
\left({\cal D}_3 v^{(k)}\right)^{c}(t)\;
S_{\alpha_1\alpha_2\alpha_3}
\label{eqn:baryon_op}
\end{equation}
can be used for all flavors of baryon and quark masses with the same displacements on the same ensemble,
and the ``perambulator''
\begin{equation}
  \tau_{\alpha\beta }(t',t) = V^\dagger(t') M^{-1}_{\alpha\beta}(t',t) V(t)
\label{eqn:peram}
\end{equation}
can be reused for different baryon (and meson) operators after a single inversion of the $M$ matrix. There is a large factor of computational power saved by ``factorizing'' correlators in terms of elementals and perambulators, and we can use the same elementals and perambulators to contract various different correlators.

\begin{figure}[t]
\begin{center}
\includegraphics[width=.32\textwidth]{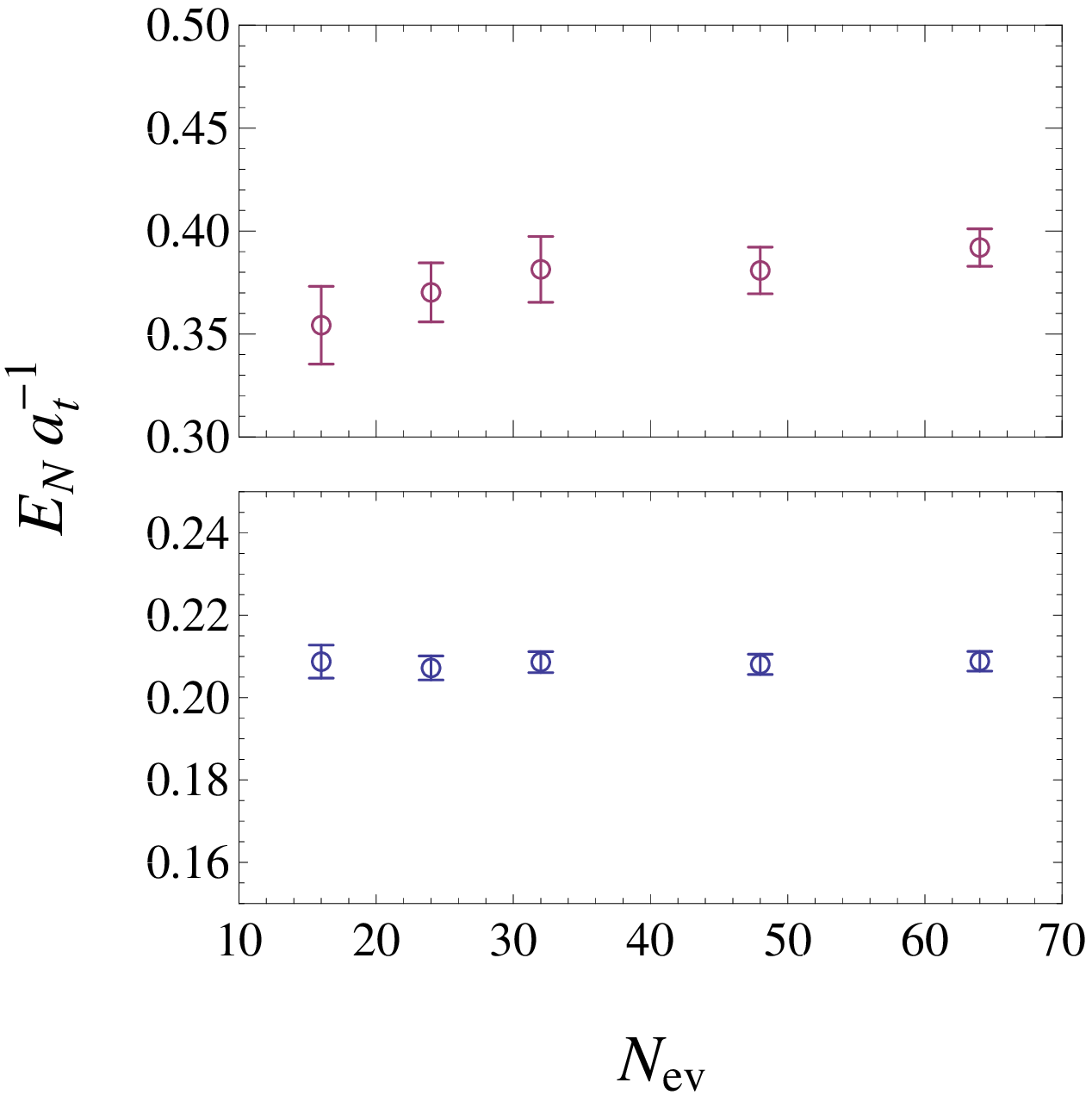}
\includegraphics[width=.32\textwidth]{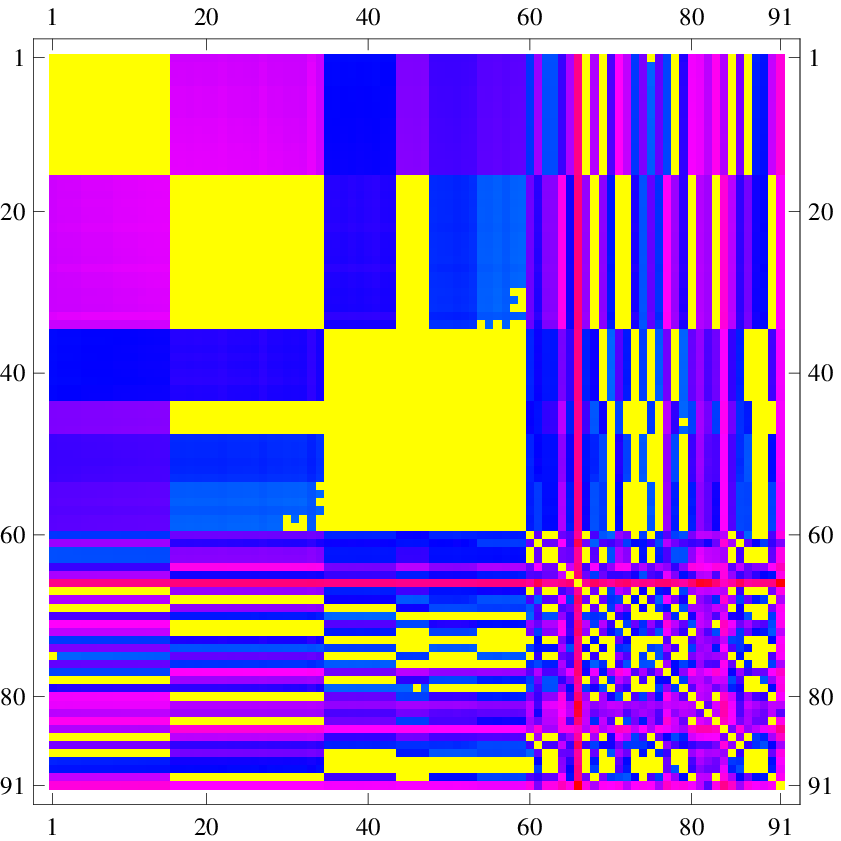}
\includegraphics[width=.34\textwidth]{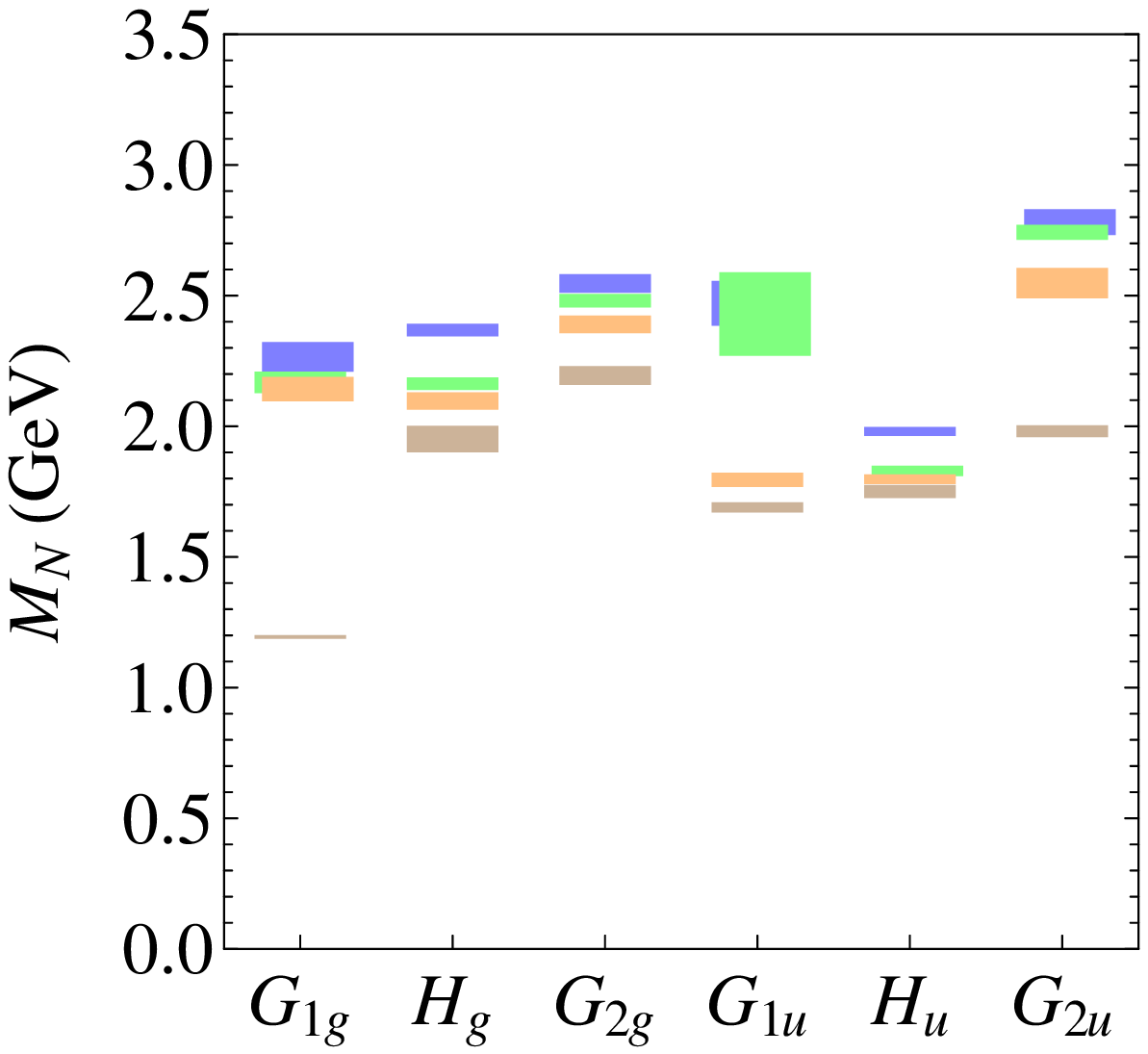}
\caption{\label{fig:spec-840}(Left) The fitted energies for the first-excited (top) and ground state (bottom) as functions of the number of distillation eigenvectors $N$. %
(Middle) A subset of $G_{1g}$ irrep operator correlators, grouped by their inner product. The yellow blocks indicate overlap $>70$\%. $G_{1g}$ is the worst case among all the irreps.
(Right) Nucleon excited spectrum sorted according to cubic-group irrep.
}
\end{center}
\end{figure}

The distillation method is demonstrated on four nucleon $G_{1g}$ operators, two of which are local to a single site and the remaining two having a singly displaced quark field. The corresponding $4 \times 4$ matrix of correlators was computed on 316 configurations of the $16^3 \times 128$ ($\approx 380$~MeV) lattice ensemble.
Modeling the correlator noise-to-signal ratio with $ a + \frac{b}{N^p}$ as a function of the number of distillation eigenvectors included ($N$ in Eq.~\ref{eqn:box}) gives a best-fit exponent $p \sim 1.1(2)$; that is, increasing the number of vectors decreases the noise considerably faster than simple statistical scaling. The variational method~\cite{VM} 
is used to extract the masses of the lowest two states in the $G_{1g}$ spectrum. The extracted mass dependence on the number of eigenvectors is shown in the left panel of Fig.~\ref{fig:spec-840}. We find consistent masses among the larger numbers of eigenvectors, with an increase in statistical precision as the number of vectors is increased.

A procedure called ``pruning'' is used to reduce the number of operators included. A practical procedure is to calculate the correlators with the same source and sink operator (i.e. the diagonal elements of the full correlator matrix) and sort them according to their ``individuality''. One way to systematically prune is to take the matrix of inner products of the effective masses of each correlator across all time slices and sort them from there. The middle panel of Fig.~\ref{fig:spec-840} shows a subset of various $G_{1g}$ displacement-operator correlators sorted by the values of their inner products with respect to one another. Any overlap larger than 70\% is marked by yellow, while the remaining values are depicted as magenta to blue colors. ($G_{1g}$ has the largest overlap amongst its operators of all the baryon irreps.) Notice that the inner-product matrix has a clear block structure when organized this way. Each yellow block consists of a set of operators that yield nearly identical effective masses as functions of time. We exclude many operators by such a selection process and pick a total of 24 operators among different quark orientations for each irrep. Then we refine to a smaller sub-matrix of $8\times 8$ by filtering the matrix by condition numbers; we perform a variational-method analysis on the resulting $8\times 8$ matrix and retain the lowest 4 eigenstates.


Fig.~\ref{fig:spec-840} shows preliminary results for nucleon spectroscopy with pion mass around 380~MeV on a $16^3\times 128$ volume, using
``distillation''~\cite{Peardon:2009gh} with $N=32$ eigenvectors. We observe a similar distribution of states as the previous study. Further measurements on the larger volume and investigation of decay thresholds and potential two-particle states are underway. (See the proceeding by K.~Juge, reporting two-particle results using distillation.)

\section{Conclusion and Outlook}
\vspace{-0.3cm}
We report an ongoing effort to solve the mysteries of baryon resonances. The  ground-state baryon masses are in reasonable agreement with experiment and consistent  among different groups with different actions if systematics are taken into consideration. Our new technique, ``distillation'', will greatly improve precision in our future calculations for extracting excited-state masses using cubic group-irrep operators, which provide powerful probes to extract highly excited resonances. A preliminary result for nucleons on $N_f= 2+1$ $m_\pi=380$~MeV is shown in this proceeding. Work on larger volumes (with a modified stochastic distillation) are under development. Meanwhile, parallel work from the HSC for meson spectroscopy with exotic quantum numbers and baryons using derivative operators are also in progress. Multi-particle operators are under investigation to distinguish these from resonances. We are also investigating the application of the distillation method to form factors to help us understand the nature of specific states.

\vspace{-0.3cm}

\end{document}